# EXPERIMENTAL INVESTIGATION OF A TEMPERATURE SEPARATION EFFECT INSIDE A SHORT VORTEX CHAMBER

Yan Beliavsky

Super Fine Ltd.
P.O.B. 532, Industrial Area, Yoqneam 20692, Israel,
E-mail: superfin@netvision.net.il

## ABSTRACT

Room temperature compressed air was pumped into a short (H/D = 0.18) vortex chamber. Experiments revealed temperature separation. The highest temperature of the periphery was 465 °C, and the lowest temperature of the central zone was -45 °C. During the process, heat was transferred to periphery in the opposite direction of the powerful vortex flow. This heat transfer cannot be explained by conventional heat transfer processes. To explain this phenomenon, the concept of Pressure Gradient Waves (PGW) is proposed in the paper. PGW are elastic waves which operate in compressible fluids with pressure gradients and density fluctuations.

## NOMENCLATURE
D     vortex chamber diameter, mm
d     outlet diaphragm diameter, mm
H     vortex chamber height, mm
h     distance from low disk to central rod, mm
P     pressure (gauge pressure), bar
T     temperature, °C
*Subscripts*
PGW     Pressure Gradient Wave
exp     Experimental

## INTRODUCTION

Several theories (from Fulton's theory [1] to current approach [2]) have been proposed to explain the effect of temperature separation inside vortex tubes (Ranque effect) [3]. These theories are based upon heat transfer processes which transfer heat from the axis area to the periphery. Such heat transfer seems quite possible inside the vortex tubes with a ratio of L/D = of 10/1 to 30/1. But current critical review [4], emphasized that future research will benefit.

Goldshtik [5] observed Ranque's effect at the outlet diaphragm of a short vortex chamber. However, he only examined the central area of the vortex chamber, were the air flow is similar to the vortex tube flow.

In this paper, I present results of experiments which reveal temperature separation inside a short (H/D=0.18) vortex chamber - with heating of the periphery and cooling of the central area. A powerful air flow moves from the periphery to the center of the vortex installation. Yet, since there is no gas displacement to the periphery any movement of the micro volume is impossible. Thus, this heat transfer process warrants further investigation.

## THE EXPERIMENT

### Experimental Equipment
The schematic experimental setup used is shown in Figure 1.

The vortex chamber consisted of lower disc (1), a cylindrical side wall (2), and an upper disc (3). On the cylindrical side wall (2) four 5mm tangential nozzles vents (7) were spaced at 90°. The air jets enter the chamber at a 20° angle to tangent line. Rotating air moved from the periphery to the center, then through a diaphragm outlet (4) to the discharge collector (5), and finally escaped through the outlet connection (6). The central rod (8) was mounted coaxial to the outlet diaphragm (4). The plugged branch pipe (9) was mounted on the outer side of the cylindrical side wall (2). The cylindrical cavity of pipe (9) was 8 mm (diameter) and 32 mm long. The installation was equipped with thermocouples and pressure transmitters. The $P_{exp}$ (bar) pressure, measured by transmitters, was a gauge pressure (atmospheric pressure = 0). These experiments were repeated with outlet diaphragms (4) of various diameters (20mm, 25mm, 30mm, 35mm, and 40mm).

### Characteristics of Sensors and Their Accuracy
The vortex chamber was supplied with compressed air having a dew point of + 3° C at temperatures from 14°C to 28°C. The inlet pressure was adjusted by a pressure regulator. The pressure readings were measured by transmitters (Model: MRT221, accuracy 0.05% F.S.). The transmitters were connected by flexible polyethylene tubes with measurement points.
The temperature readings were measured by thermocouples (RTD type, accuracy ±0.2° C and K type, accuracy ±0.5°C). The thermocouple junction diameters were 1.0÷1.5 mm. The thermocouples were thermo-isolated from stainless steel parts by Teflon pieces (d = 10 mm).

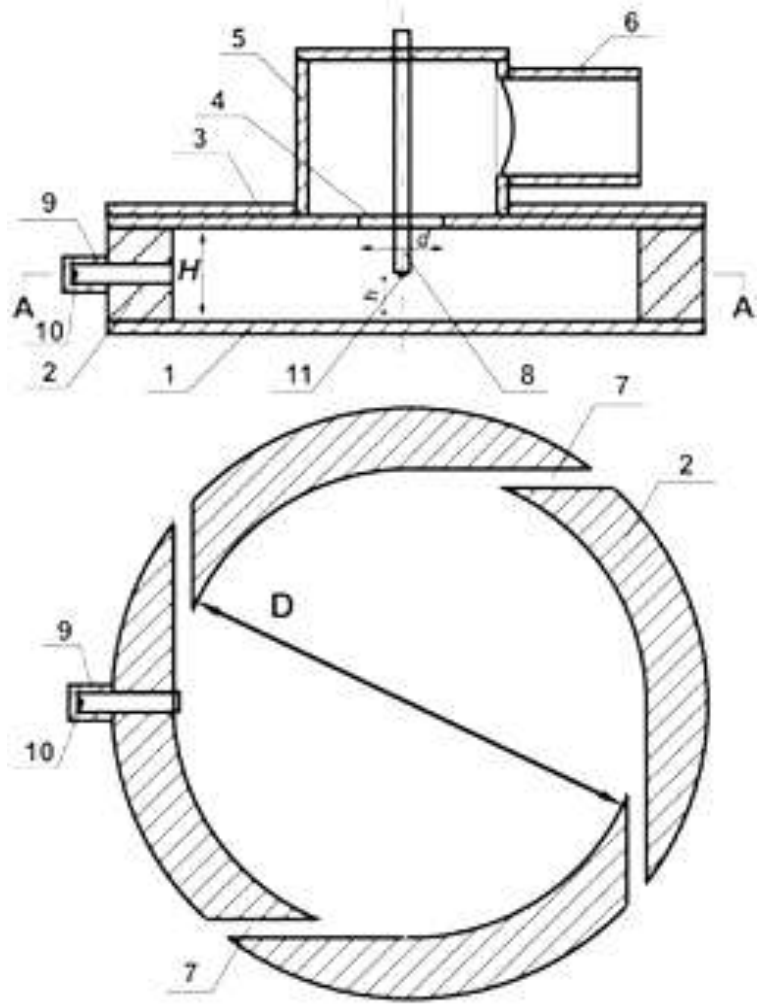

**Figure 1** Schematic overview of experimental vortex chamber; Top: cross-section front view; Bottom: cross-section top view. The experimental setup stand of the following components: 1. Lower disc, 2. Cylindrical side wall, 3. Upper disc, 4. Outlet diaphragm, 5. Discharge collector, 6. Outlet connection, 7. Tangential nozzles, 8. Central rod, 9. Plugged branch pipe, 10. "Hot" thermocouple, 11. "Cold" thermocouple. $D$ – Vortex chamber diameter (140 mm); $H$ - Vortex chamber height (25 mm); $d$ - Outlet diaphragm diameter; $h$ – Distance between Central rod and Lower disc.

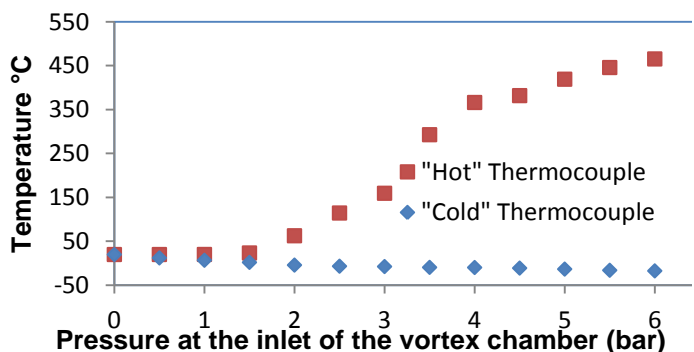

**Figure 2** The temperature (°C) inside the branch pipe as a function of the inlet pressure (bar):
The experimental setup with outlet diaphragm $d$=30mm.

### Results

The characteristics of vortex flow measured inside the chamber (Figure 1) are listed in the below.

Figure 2 shows the relationship between temperatures (measured by two thermocouples Figure 1 and inlet pressures for the outlet diaphragm $d$ = 30 mm. One thermocouple (11) was mounted at lower end of the central rod, and the other thermocouple (10) was mounted at plugged end-cap of branch pipe.

The temperatures for outlet diaphragm diameters of 20 mm, 25 mm, 35 mm, and 40 mm are shown at Figure 3. Notice that for the $d$ = 30 mm outlet diaphragm the heating effect (with a maximal value of temperature of 465°C at a pressure of $P_{exp}$ = 6 bar), is much stronger

1483

than for the other diameter sizes. The cooling effect of the "cold" thermocouple is relatively similar.

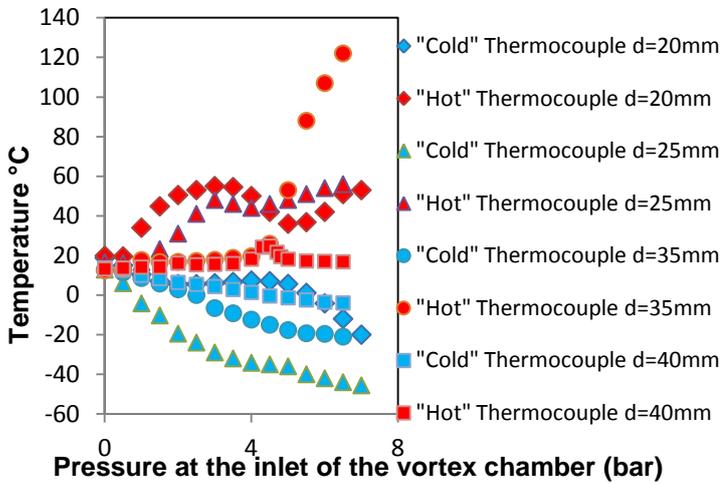

**Figure 3** The temperature (°C) inside the branch pipe as a function of the inlet pressure (bar):
The experimental setup with outlet diaphragms $d =$ (20mm, 25mm, 35mm and 40mm).

The minimal cold temperature -45°C for the $d = 25$ mm was fixed at an inlet pressure of Pexp = 7.0 bar. The function of the cooling of pressure for diaphragms (20 mm, 25 mm, 35 mm, and 40 mm) was monotonic; while the heating function had local maxima and minima.

The level of heating/cooling depends on the symmetry of the assembly. The position of central rod axis was displaced from common axis. For the displacement magnitudes (0; 1mm; 3mm, and 7mm) "hot" thermocouple (10) temperatures of (170°C; 150°C; 60°C and 38°C) were obtained, respectively ($d = 30$ mm; inlet pressure Pexp = 4.0 bar; inlet temperature 18°C).

The temperature measured by "hot" thermocouple (10) and the side wall pressure value also depended on the distance $h$ between the lower end of the central rod and the lower disk.
Figure 4 shows the experimental temperature variation of the "hot" thermocouple (10). Distance $h$ was changing from the lower end of the central rod to the lower disk. The experiment was performed with an outlet diaphragm of $d = 30$ mm at constant inlet pressure Pexp = 4.0 bar. The first measurement was $h = 0$, the central rod touched the lower disk. The rod was kept at 5mm distances until the thermocouple reading stabilized and moved 5mm further. Finally, at $h = 110$ mm the central rod was completely removed from the hole on the lid of the discharge collector. In this position, thermocouple heating stopped completely. The temperature of 37°C was not final, but corresponds to a ~ 10 minute cooling time (not the stabilization time). The maximum temperature was 212°C at the $h = 50$ mm. The function (Figure 4) has local maximum and minimum.
The position of central rod also affected on the side wall pressure value. Figure 4 shows the side wall pressure as function of the rod position. The pressure values were averaged at the appropriate rod position for the data recorded. The pressure curve inversely corresponds to the temperature curve; the higher the temperature inside the branch pipe, the lower the pressure on the side wall.
During the heating of the branch pipe, the side wall and lower disc were also heated, but not to a high temperature. Towards the center, the temperatures gradually decreased.

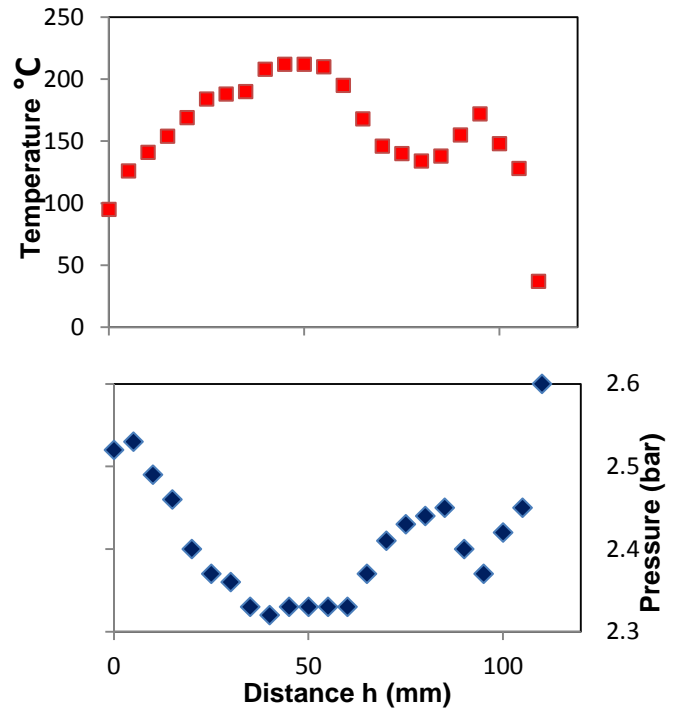

**Figure 4** Temperature (°C) measured by "hot" thermocouple placed at plugged end-cap of branch pipe and the pressure (bar) on the side wall as functions of the position of central rod (mm):

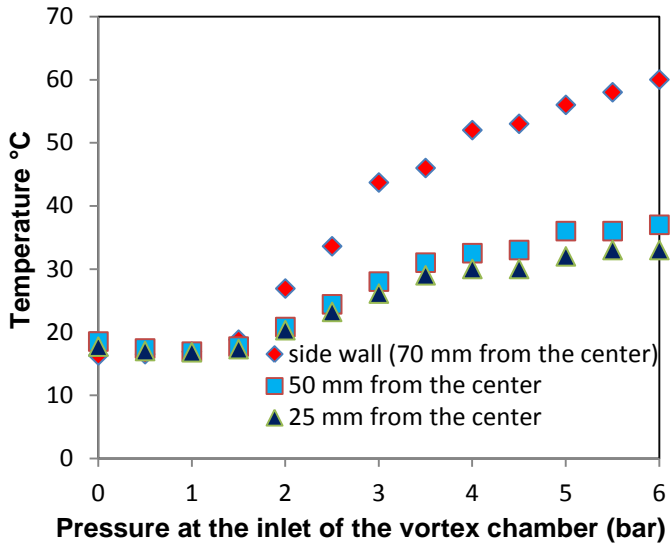

**Figure 5** The temperatures (°C) inside the vortex chamber as a function of the inlet pressure (bar) for the experimental setup with outlet diaphragm $d$ = 30 mm. The thermocouples were mounted at the points: side wall (70 mm from the center of vortex chamber) and low disc (50 mm and 25mm from the center).

Figure 5 shows that the reading of thermocouples depended on input pressure. Figure 5 and Figure 2 show the temperature readings obtained during same experiment with $d$ = 30 mm and $h$ = 15 mm. For example, at an inlet pressure of $P_{exp}$ = 6.0 (bar) the following temperatures values were reached: on the side wall 60°C; at a point 50 mm from the center 37°C; and at a point 25mm from the center 33°C (inlet air temperature was 18°C).

During maximal heating ($d$ = 30 mm), large amounts of heat were released on the periphery of the vortex chamber. Not only the branch pipe warmed up, but the entire massive side wall was heated. The branch pipe, made of PVC, melted and severed by the pressure. The polyethylene tubes, connecting the pressure transducers with the side wall and lower disc, were also broken by the heat and pressure.

With the introduction of air, the process was accompanied by much loud noise. The analysis of the audio spectrum, executed with $d$ = 30 mm, shows a distinct narrow peak (~ 3.000 Hz) at input pressures above 1.7 bar. The amplitude of the peak grew with increasing pressure and considerably more than the baseline noise level. Pressure increases produced a distinct peak at ~ 6.000 Hz. Preliminary observations (to be further examined) indicate that the effect of temperature separation correlates with the level of "loudness".

**Vortex Flow Characteristics**

The chart Figure 6 shows the measured values of the pressures inside vortex chamber (Figure 1) at an input pressure 6.0 bar and a distance of $h$ = 15mm.

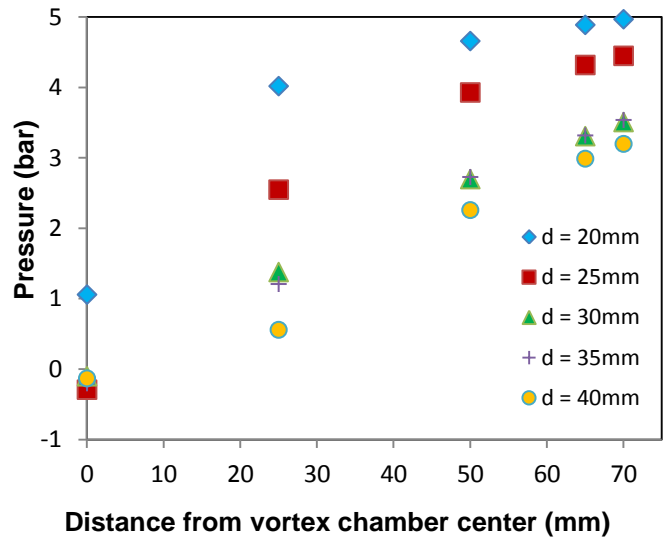

**Figure 6** Pressures inside vortex chamber (bar) at a corresponded distance (mm) from the center. The measured values are plotted for sets with different outlet diaphragm diameters: 20mm, 25mm, 30mm, 35mm and 40mm.

Experimental points were plotted on the radius of the chamber for sets with different diaphragm diameters $d$. Distance from the center ($r$ = 70 mm) represents a point at the side wall. For example, the value of pressure on the side wall for diaphragm $d$ = 30 mm is equal to 3.5bar. The vortex flow generates a considerable pressure gradient for all diameters of outlet diaphragms. The smaller the diaphragm diameter $d$ is, the higher the pressure values.

The pressure at the axis of the vortex chamber (point 0) for diaphragm $d$ = 20 mm is equal to $P_{exp} \cong$ 1bar, i.e. the flow regime is "submerged". It should be noted that a 16 mm central rod was mounted on the axis. The air flow rate passed through a ring slot of 2 mm thick. For the other diaphragm diameters (25mm, 30mm, 35mm and 40mm), the value of pressure at the center of the chamber (point 0) was negative.

The chart (Figure 7) shows the ratio $\alpha$ as a function on the inlet pressure of the vortex chamber. Measurements were recorded for sets with different diaphragm diameters ($d$) and distance ($h$ = 15 mm). The Pinput and Pside wall were the pressures (bar) at the inlet of vortex chamber and at the side wall, respectively.

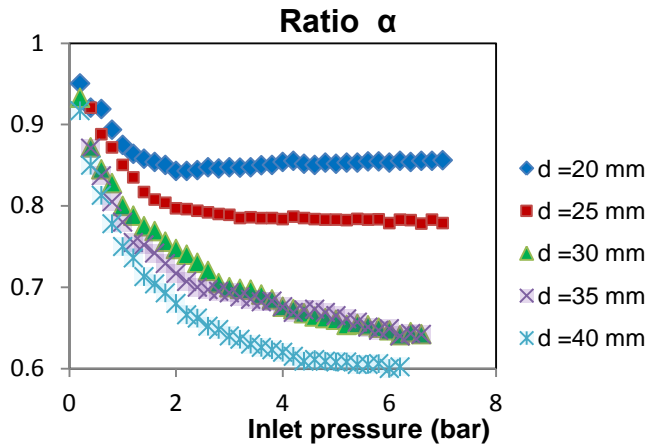

**Figure 7** Ratio **α** as function on pressure at the inlet of vortex chamber (bar). The values $\alpha$ were plotted for sets with different diameters $d$ of outlet diaphragm: 20mm, 25mm, 30mm, 35mm and 40mm.

$$\alpha = \frac{P\text{side wall} + 1}{P\text{input} + 1}$$

The value of ratio $\alpha$ established the magnitude of velocity of air jets, entering the vortex chamber. Notice that the value $\alpha$ is greater than 0.53 (critical value for air) for all diaphragms in all ranges of input pressure. This means that the jet velocities at the input to the vortex chamber are less than the theoretical sound velocity. The real velocities were even less, correcting the friction loss inside the nozzles. Therefore, the vortex flow inside the chamber was not accompanied by shock wave propagation.

While the vortex flow moved from periphery to the center, the radial and tangential component velocity increased. Thus, due to the air expansion, the increase was steeper than hyperbolically (as for incompressible fluid).

The Table 1 represents the results of measurements indicate the rotation velocity inside the vortex chamber.

**Table 1**. Impeller rotation as function of inlet pressure

| The pressure at the inlet of vortex chamber (bar) | 0.5 | 1 | 2 | 3 | 4 | 5 |
|---|---|---|---|---|---|---|
| Rotation speed (rpm) $\times 10^3$ | 14 | 18 | 20 | 28 | 34 | 35 |

An impeller was mounted at the vortex chamber axis (d = 90mm and a height of 20mm, with four radial blades). The real air flow rotation speed exceeded the values presented in Table 1. The friction heated the bearing pair to significant temperatures, especially at high pressures.

**Discussion**

If inside the vortex tubes [3], heat is transferred from the center to periphery by "hydrodynamics" (diffusion + convection) [2], the thermal separation process, observed during the experiments and the Ranque effect are two distinct processes. If the "hot" micro-volumes are present inside the vortex chamber Figure 1, they cannot move to periphery by the powerful radial flow towards the center. So, "hydrodynamic" heat transfer from center to periphery is impossible.

The author cannot suggest an adequate description of the results (Figure 4) (based on the hydrodynamic performances) how the distance $h$ affects the temperature inside the branch pipe. Note that the central rod movement above $h = 25$ mm is out of the vortex chamber and inside the discharge collector.

It is reasonable to assume that the dependence of side wall pressure on the central rod position (Figure 4) was defined by the temperature of the wall. The wall temperature changed with the readings of the "hot" thermocouple inside the branch pipe. This temperature corresponded to a temperature of wall layer of the vortex flow. Since the supply of heat to a subsonic gas flow inside a tube led to a flow velocity increase concurrent with pressure reduction, probably a similar process occurred in the vortex flow. Apparently, the wall layer of the vortex flow is under the same "space-limited" conditions. Such an assumption corresponds to the experimental results presented. With the increase of side wall temperature, the wall layer pressure decreased indicating that increasing the heat to vortex periphery increases its kinetic energy.

Furthermore, the vortex flow inside the above vortex chamber excludes the formation of shock waves.

In addition, in the absence of other heat sources, the possibility of heat transfer to periphery by radiation inside this apparatus need not been considered.

**Consequently, the author cannot explain the results of these temperature experiments, based on the "hydrodynamic" principles.**

**PRESSURE GRADIENT WAVE**

An alternative explanation is a Hypothesis based upon the possible existence of a special kind of elastic waves that occur in compressible media with pressure gradient.

Assume the sound fluctuations of density split the space (equal to the wave length) in two zones - a higher density and a lower density zone. With a pressure gradient, resultant forces emerge in the areas of the fluctuations. In such circumstances, a secondary wave (Pressure Gradient Wave) (PGW) may have occurred. In this case, these waves move along the vector of the pressure gradient and include two wave fronts: a "sinking" wave of the compression directed toward the higher pressure, and the "floating" wave of expansion, in the direction of lower pressure. Concurrently, the compressed wave carries increased pressure and temperature, and the rarefied wave has reduced pressure and temperature.

Although, the PGW concept still requires experimental evidence and serious theoretical development, we can note some of its properties. PGW increases the temperature in the high pressure zone and reduces it in the low pressure area. PGW transfer heat from a low to a high pressure zone using the energy of the pressure gradient, without a temperature gradient. The sound resonance is an important factor in PGW emergence. Resonance leads to the increasing of initiating density fluctuations.

The presented experimental results can be explained by such a wave theory. Rotation of air vortex creates the pressure gradient. The configuration of the vortex chamber with air flow features affect the frequency and amplitude of generated sound waves. The outlet diaphragm diameter d=30 mm forms a sound frequency close to the natural resonance frequency of the vortex chamber (3.000 Hz is the main frequency; 6.000 Hz is the first harmonic). At this diameter, optimal resonance conditions were observed. The amplitude of density fluctuations is maximal that creates a powerful PGW. Other diaphragm diameters and changing the central rod position change the resonance condition. The amplitude of initiating density fluctuations decrease and the PGW will also decrease or not exist.

The hypothesis of Pressure Gradient Waves could relate both to the vortex chamber described in this paper and vortex tubes [3] in general. PGW are produced by the vortex flow rotation and sound inside vortex tubes. Sprenger [6] has shown that standing sound waves exist inside vortex tubes.

Perhaps, PGW also operate inside Hartmann sound generators [7]. In these devises, the pressure gradient is created by gas jet impact. PGW heat the end of the cavity, mounted opposite to the nozzle. During an experiment [8], using helium jet, the end cavity temperature reached ~1000°C. On the basis of the PGW phenomenon, new heat transfer devices (including heat pumps) which operate in any temperature diapasons could be developed [9].

## CONCLUSIONS

The experiment presented confirms that the temperature separation phenomenon occurs inside the vortex chamber with the heating of the periphery and the cooling of the central zone. The highest temperature of the periphery was 465 °C, and the lowest temperature of the central zone was -45 °C. During the temperature separation process, heat was transferred in the opposite direction of the powerful vortex flow. The author cannot explain the results of experiments based on the "hydrodynamic" performances.

The concept of Pressure Gradient Wave (PGW), new kind of elastic waves which transfer heat is proposed. PGW always propagate along the pressure gradient vector. The process is triggered by sound. PGW transfer energy from the low-pressure to the high-pressure zones. To strengthen PGW, resonance conditions must exist inside the device. The concept of PGW still requires experimental evidence and serious theoretical development.